# Concentration ratio for a circular trough with flat receiver

Matteo Timpano, Thomas A. Cooper*

Department of Mechanical Engineering, York University, Toronto, Canada
*Corresponding author: tcooper@yorku.ca

**Abstract**

We present a simple analytical formula for the geometric concentration ratio of a circular trough optical concentrator coupled to a flat (planar) receiver. The receiver is sized to achieve full collection (no spillage) for any combination of acceptance and rim angle. The development reveals the existence of two concentration ratio regimes, one controlled by the edge ray caustics, and the other controlled by the intersection of the edge rays from the rim alone. Notably, we show that in the latter rim ray regime, the circular trough achieves the exact same geometric concentration ratio as a parabolic trough. The provided formula fills a gap towards the development of a complete catalogue of concentration ratio formulas for different concentrator/receiver geometries.

## 1 Introduction

When designing a concentrating solar collector, one of the first steps is to size the concentrator and receiver. This sizing is guided by the achievable geometric concentration ratio $C_g$ – defined as the ratio of the collecting area $A_i$ of the concentrator to the absorbing area of the receiver $A_o$ – for a given concentrator and receiver geometry. To this end, simple closed-form expressions for the achievable concentration ratio for common concentrator/receiver geometries are of considerable utility to the optical designer. Such concentration ratio formulas exist for several common geometries, e.g. parabolic troughs with flat and circular receivers [1], asymmetric parabolic troughs [2], and compound parabolic concentrators [1, 3, 4]. Surprisingly, there are several potentially useful concentrator/receiver geometries, most notably circular troughs, for which no simple concentration ratio formula has been presented in the literature. It is the view of the authors that a complete catalogue of concentration ratio formulas would be of significant archival value to the solar energy community to enable a quick comparison of different designs. In this short communication, we therefore present the development of the concentration formula for a concentrator/receiver geometry of potential practical interest, namely a circular trough concentrator with a flat (planar) receiver. This work builds upon the work of [5], which interestingly focused on triangular receivers for circular mirrors, and the work of [6] where a partial solution for a flat absorber, without derivation, was presented.

Our development reveals several important features of the circular trough concentrator, most importantly the existence of two distinct concentration regimes, one governed by the caustics and one governed solely by the intersection of the edge rays from at the rim. Remarkably, we find that in the rim ray regime, the concentration ratio achieved by a circular trough with plane receiver is identical to that of a parabolic trough. This result is of practical significance since a circular profile can be constructed with relative ease compared to a parabolic profile, for example by inflating a partially mirrored polymer film cylinder with a modest overpressure [5, 7–10].

## 2   Geometry and Definitions

**Figure 1** shows the geometry of a circular trough concentrator whose cross-section comprises a circular arc symmetric about the $z$-axis (optical axis), coupled to a flat receiver. With the center of the circular arc is placed at the origin **O**, the mirror may be parameterized by

$$\mathbf{P} = \begin{bmatrix} P_x \\ P_z \end{bmatrix} = \begin{bmatrix} R\sin\omega \\ -R\cos\omega \end{bmatrix} \quad (1)$$

where $R$ is the radius of the circle and $\omega$ is the parametric angle measured positive counterclockwise from $-z$, and spanning from the left rim (LR) at $-\omega_{rim}$ to the right rim (RR) at $\omega_{rim}$.

Consider a ray incident on any point **P** of the mirror. An "on-axis" ray, i.e. any incident ray parallel to the $z$-axis, will be reflected such that it makes an angle $\varphi = 2\omega$ with the $z$-axis. The polar angle $\varphi$ that an on-axis reflected ray makes with the optical axis is a useful quantity for comparison to other concentrators. The polar angle at the rim $\varphi_{rim} = 2\omega_{rim}$, commonly referred to as the "rim angle", represents the maximum angle that a reflected on-axis ray makes with respect to the optical axis as it approaches the focus.

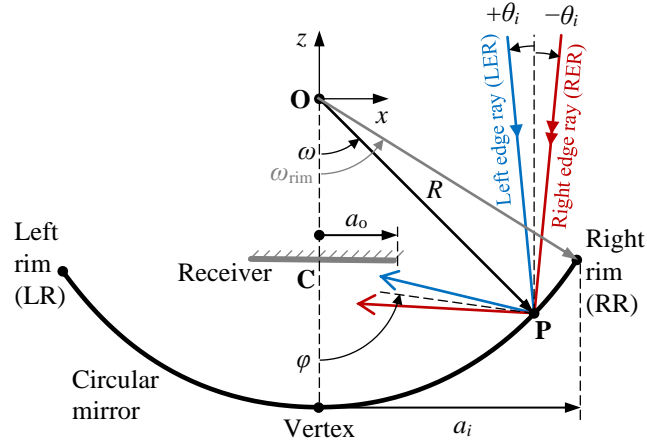

**Figure 1.** Geometry of a circular trough concentrator coupled to a flat (planar) receiver, showing the main parameters required to determine the achievable concentration ratio $C_g = a_i/a_o - 1$.

In general, rays will be incident from within an "acceptance cone" of $\pm\theta_i$ where $\theta_i$ is the acceptance (half) angle of the concentrator. This acceptance angle will typically be sized to accommodate all angular errors in the system including the angular size of the sun/sunshape [11], slope/shape/specularity error [12], tracking error, and skew dilation [6, 13]. A value of 0.5° to 1° [1] is typical of continuously tracking systems, while much larger angles may be used for stationary or seasonally adjustable designs [1, 5, 14]. We refer to any ray incident at an angle $+\theta_i$ as a left edge ray (LER), and any ray incident at an angle $-\theta_i$ as a right edge ray (RER).

The direction vector of any ray reflected by the mirror may be parameterized by

$$\mathbf{v} = \begin{bmatrix} v_x \\ v_z \end{bmatrix} = \begin{bmatrix} -\sin(\varphi - \theta) \\ \cos(\varphi - \theta) \end{bmatrix} = \begin{bmatrix} -\sin(2\omega - \theta) \\ \cos(2\omega - \theta) \end{bmatrix} \quad (2)$$

where $\theta$ is the angle of the *incident* ray measured counterclockwise from the $+z$ axis. A generic reflected ray may then be parameterized by

$$\vec{\mathbf{r}} = \begin{bmatrix} r_x \\ r_z \end{bmatrix} = \mathbf{P} + t\mathbf{v} = \begin{bmatrix} R\sin\omega - t\sin(2\omega - \theta) \\ -R\cos\omega + t\cos(2\omega - \theta) \end{bmatrix} \quad (3)$$

where $t$ is the pathlength traveled by the ray after reflection.

The receiver is to be positioned with its center at some point $\mathbf{C}$ along the optical axis, and we seek the smallest receiver size that intercepts *all* rays reflected from the mirror within $\pm\omega_{rim}$. A convenient approach to size the receiver such that all rays are intercepted is to make use of the caustic of the edge rays [5, 15]. For the geometry shown in **Figure 1**, the caustic curve is found from the solution of [15]

$$\frac{\partial r_x}{\partial t}\frac{\partial r_z}{\partial \omega} - \frac{\partial r_x}{\partial \omega}\frac{\partial r_z}{\partial t} = 0 \qquad (4)$$

Subbing in Eq. (3) yields

$$2t - R\cos(\omega - \theta) = 0 \qquad (5)$$

and solving for $t$ yields

$$t = \tfrac{1}{2} R\cos(\omega - \theta) \qquad (6)$$

Subbing Eq. (6) into Eq. (3) yields the caustic for a circular mirror, which exhibits the well-known nephroid form [16, 17]

$$\mathbf{c} = \begin{bmatrix} c_x \\ c_z \end{bmatrix} = \begin{bmatrix} \tfrac{1}{4} R(3\sin\omega - \sin(3\omega - 2\theta)) \\ \tfrac{1}{4} R(-3\cos\omega + \cos(3\omega - 2\theta)) \end{bmatrix} \qquad (7)$$

Equation (7) may be used to construct the caustic produced by any set of rays incident at a given angle $\theta$ with respect to the optical axis. Using Eq. (3) and (7) it is now possible to size the receiver to intercept all rays reflected by the mirror.

## 3 Derivation of the concentration ratio

We wish to size the receiver for "full collection" meaning that it intercepts all radiation incident within the acceptance cone $\pm\theta_i$ that is reflected by the primary over its full span from $-\omega_{\text{rim}}$ to $\omega_{\text{rim}}$. We seek the position $C_z$ and width $2a_o$ of the smallest possible receiver that is just large enough to achieve full collection. A receiver thusly sized will achieve the highest possible full-collection geometric concentration ratio, defined as

$$C_g \equiv \frac{A_i}{A_o} = \frac{2a_i - 2a_o}{2a_o} = \frac{a_i}{a_o} - 1 \qquad (8)$$

which we refer to as the "full collection concentration limit". The $-1$ term accounts for self-shading of the concentrator inlet aperture by the receiver, implying that the receiver is "one-sided" (can only receive radiation incident from the bottom). It is important to note that this concentration limit results from the particulars of a given concentrator/receiver geometry, and will always be lower than the thermodynamic limit [4]

$$C_{g,\text{ideal},2D} = \sin\theta_o / \sin\theta_i \tag{9}$$

where $\theta_o$ is the half-angular range of the beam at the receiver ($\theta_o = \varphi_{\text{rim}} + \theta_i$ for this geometry). Using Eq. (8), we may develop the concentration formula by finding the values of $a_i$ and $a_o$ as a function of the design parameters for the concentrator, namely the acceptance angle $\theta_i$ and the rim angle $\varphi_{\text{rim}}$.

The half-width of the inlet aperture follows directly from **Figure 1** and Eq. (1)

$$a_i = P_x = R\sin\omega_{\text{rim}} \tag{10}$$

We are then left to determine $a_o$, which requires consideration of the trajectories of the rays after being reflected by the mirror. Interestingly, we will show that there are two possible conditions for defining the minimum required receiver size. This results in two concentration ratio "regimes", one defined by the caustics (caustic regime), and another defined solely by the rays reflected off the rim of the concentrator (rim ray regime). The correct regime is that which yields the larger receiver size (smaller concentration ratio) for a given acceptance and rim angle, as the smaller receiver from the other regime would necessarily lead to some rays missing the receiver.

*3.1 Caustic regime*

In the caustic regime, the receiver size may be determined from the caustics. **Figure 2 a)** shows the caustic formed by the right edge rays ($\mathbf{c}_{\text{RER}}$) and the caustic formed by the left edge rays ($\mathbf{c}_{\text{LER}}$). As seen in **Figure 2 a)**, the smallest possible receiver must span from: 1) the rightmost point on the left edge ray caustic $\mathbf{c}_{\text{LER}}$; to 2) the leftmost point on the right edge ray caustic $\mathbf{c}_{\text{RER}}$.

Condition 1) and 2) may be written mathematically as

$$\mathbf{c}(-\omega_{\text{rim}}, -\theta_i) \tag{11}$$
$$\mathbf{c}(\omega_{\text{rim}}, \theta_i) \tag{12}$$

respectively. Due to symmetry about $z$, only one of these points is needed. Using condition 2 (Eq. (12)), the half-width of the receiver in the caustic regime is found to be

$$a_{o,\text{caustic}} = c_x(\omega_{\text{rim}}, \theta_i) = \tfrac{1}{4} R\left(3\sin\omega_{\text{rim}} - \sin(3\omega_{\text{rim}} - 2\theta_i)\right) \tag{13}$$

## 3.2 Rim ray regime

It is observed from **Figure 2 a)** that a distinguishing feature of the caustic regime is that the right edge ray reflected off the right rim, RER-RR, (and similarly for the LER-LR) falls within the limits of the receiver automatically. However, for larger rim angles and acceptance angles, the RER-RR becomes more horizontal (rotates counterclockwise), such that it will eventually spill over the left-hand side of the receiver. At this point, it is necessary to enlarge the receiver, beyond that predicted by the caustics alone, to capture this rim ray. The receiver size can then be determined by the intersection of the edge rays from the rim alone, without the need for the caustics.

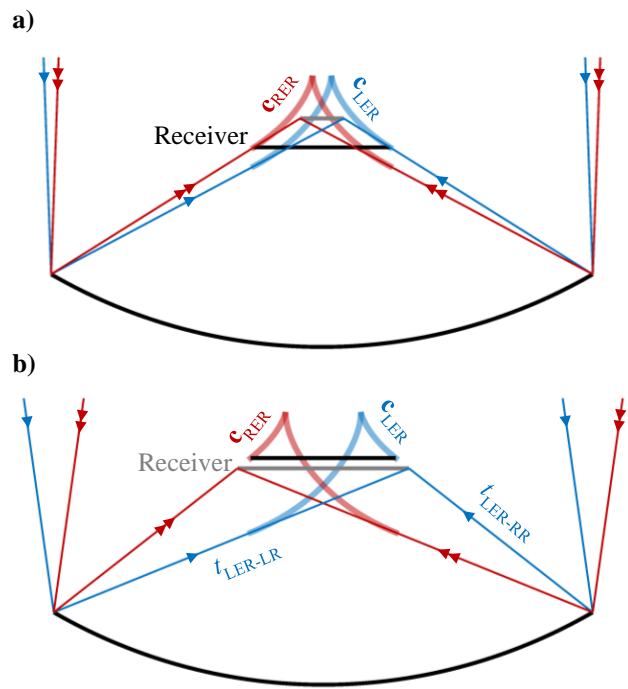

**Figure 2.** Comparison of conditions for sizing the receiver in the caustic and rim ray regimes. a) Caustic regime design. This concentrator has $\omega_{rim} = 30°$, $\theta_i = 2°$ and achieves $C_g = 2.98\times$. b) Rim ray regime design. The gray line is the receiver, which is larger than predicted by the caustics (black line). This concentrator has $\omega_{rim} = 30°$, $\theta_i = 8°$ and achieves $C_g = 2.14\times$.

**Figure 2 b)** shows a concentrator design in the rim ray regime. The smallest possible receiver must span from: 1) the intersection of the RER from the right rim and the RER from the left rim; to 2) the intersection of the LER from the left rim and the LER from the right rim. Due to symmetry, only one of these conditions is needed to size the receiver. Proceeding with condition 2), we may express this condition mathematically as

$$\vec{r}(+\omega_{rim},+\theta_i) = \vec{r}(-\omega_{rim},+\theta_i) \tag{14}$$

Using Eq. (3) we find

$$
\begin{aligned}
x: &\quad R\sin\omega_{rim} - t_{LER\text{-}RR}\sin(2\omega_{rim}-\theta_i) = -R\sin\omega_{rim} - t_{LER\text{-}LR}\sin(-2\omega_{rim}-\theta_i) \\
z: &\quad -R\cos\omega_{rim} + t_{LER\text{-}RR}\cos(2\omega_{rim}-\theta_i) = -R\cos\omega_{rim} + t_{LER\text{-}LR}\cos(-2\omega_{rim}-\theta_i)
\end{aligned}
\quad (15)
$$

From the $x$-equation we find

$$
t_{LER\text{-}LR} = \frac{-2R\sin\omega_{rim} + t_{LER\text{-}RR}\sin(2\omega_{rim}-\theta_i)}{\sin(-2\omega_{rim}-\theta_i)}
\quad (16)
$$

Subbing Eq. (16) into the $z$-component of Eq. (15), and solving for $t_1$, we find

$$
t_{LER\text{-}RR} = \frac{2R\sin\omega_{rim}}{\sin(2\omega_{rim}-\theta_i) - \cos(2\omega_{rim}-\theta_i)\tan(-2\omega_{rim}-\theta_i)}
\quad (17)
$$

Now we can evaluate the $x$-coordinate of the intersection point

$$
\begin{aligned}
a_{o,\text{rim ray}} &= r_x(+\omega_{rim},+\theta_i) = R\sin\omega_{rim} - t_{LER\text{-}RR}\sin(2\omega_{rim}-\theta_i) \\
&= R\sin\omega_{rim} - \frac{2R\sin\omega_{rim}\sin(2\omega_{rim}-\theta_i)}{\sin(2\omega_{rim}-\theta_i) - \cos(2\omega_{rim}-\theta_i)\tan(-2\omega_{rim}-\theta_i)} \\
&= R\sin\omega_{rim}\frac{\sin(2\theta_i)}{\sin(4\omega_{rim})}
\end{aligned}
\quad (18)
$$

### 3.3 Final $C_g$ formula

Plugging Eqs. (10) and (13) into Eq. (8), after some simplification we find the concentration ratio for the caustic regime

$$
C_{g,\text{caustic}} = \frac{a_i}{a_{o,\text{caustic}}} - 1 = \frac{4\sin\omega_{rim}}{3\sin\omega_{rim} - \sin(3\omega_{rim}-2\theta_i)} - 1
\quad (19)
$$

And similarly using Eqs. (10), (18) and (8) for the rim ray regime

$$
C_{g,\text{rim ray}} = \frac{a_i}{a_{o,\text{rim ray}}} - 1 = \frac{\sin(4\omega_{rim})}{\sin(2\theta_i)} - 1
\quad (20)
$$

To afford comparison to existing formulas for other concentrator types, Eq. (19) and (20) can be recast in terms of the polar angle $\varphi = 2\omega$ by subbing in $\omega_{rim} = \tfrac{1}{2}\varphi_{rim}$, and combined together to find the final form of the $C_g$ formula

$$C_g = \begin{cases} \dfrac{4\sin\left(\frac{1}{2}\varphi_{\text{rim}}\right)}{3\sin\left(\frac{1}{2}\varphi_{\text{rim}}\right)-\sin\left(\frac{3}{2}\varphi_{\text{rim}}-2\theta_i\right)}-1 & \text{caustic regime} \\ \dfrac{\sin\varphi_{\text{rim}}\cos\varphi_{\text{rim}}}{\sin\theta_i\cos\theta_i}-1 & \text{rim ray regime} \end{cases} \quad (21)$$

To determine the applicable branch (regime) of Eq. (21) the minimum $C_g$ of the two regimes for a given $\varphi_{\text{rim}}$ and $\theta_i$ is taken. **Figure 3** shows the resulting concentration ratio as a function of the rim angle for a range of acceptance angles.

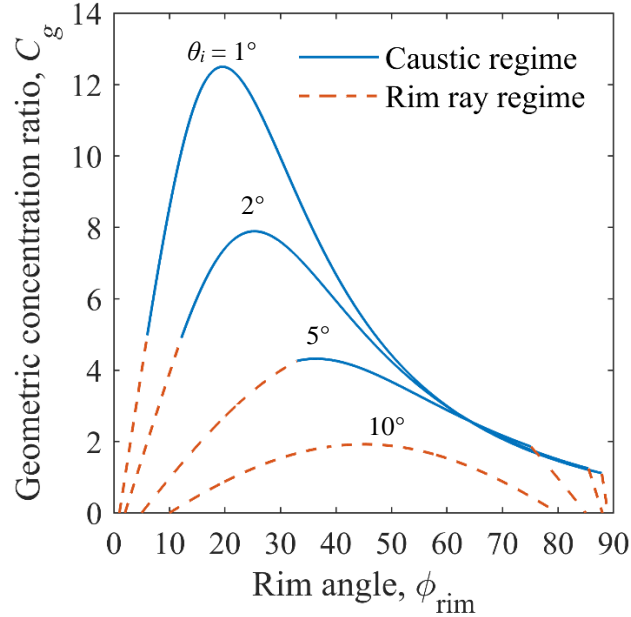

**Figure 3.** Concentration ratio vs. rim angle for a circular concentrator coupled to a one-sided flat receiver, for acceptance half-angles of 1°, 2°, 5° and 10°.

## 4 Discussion

Equation (21) is the first report of a rigorous equation for the geometric concentration ratio of a circular trough with flat receiver. Interestingly, the concentration ratio in the caustic regime was previously noted in [6], and a graphical result was presented in [5], however both without mention of the rim ray regime. Several important observations can be made by analyzing Eq. (21) and **Figure 3** in detail. First, the rim ray regime is seen to prevail for small and large rim angles, whereas the caustic regime prevails for intermediate rim angles. For larger acceptance angles, the rim ray regime governs a greater range of $\varphi_{\text{rim}}$, until at a certain value of the acceptance angle ($\theta_i = 6.6°$), the rim ray regime prevails for all rim angles. Interestingly, in the rim ray regime, the peak of the concentration ratio curve occurs for a rim angle of 45° for all acceptance angles, whereas in the caustic regime the peak occurs for smaller rim angles.

Closer examination of Eq. (21) reveals an important finding: in the rim ray regime, the concentration ratio of a circular trough with flat absorber is identical to that of a parabolic trough with flat absorber [4, 18, 19]

$$C_{g,\text{parab}} = \frac{\sin\varphi_{\text{rim}} \cos\varphi_{\text{rim}}}{\sin\theta_i \cos\theta_i} - 1 \quad (22)$$

The equal concentration of a circular and parabolic trough in the rim ray regime results from the fact that the slope at the rim is fixed by $\varphi_{\text{rim}}$. Because of this, for a given inlet aperture width, the resulting receiver size is fully determined by $\varphi_{\text{rim}}$ and $\theta_i$ without any knowledge of the mirror profile[1]. Therefore, in the rim ray regime, the parabola, the circle, or any other shape whose focal function [18] lies within the intersections of the edge rays from the rim, will yield the same geometric concentration ratio. The concentration ratio given by Eq. (22) is in fact the maximum possible full-collection concentration ratio that can be achieved by *any* single-reflection focusing mirror with a one-sided flat absorber [18].

The fact that a circular trough achieves the same concentration ratio as a parabolic trough is an important outcome, considering the potential manufacturing benefits of a circular profile. However, since the rim ray regime is primarily of importance for large values of $\theta_i$, it is anticipated that circular troughs are most appropriate for low concentration designs, e.g. seasonally adjustable designs [5, 14]. As shown by [14] an east-west axis trough with a large acceptance angle can be designed to operate year-round with only periodic adjustment, without continuous tracking. As an exemplary non-tracking design, we consider a circular trough with acceptance angle 9°, which allows for year-round operation for at least 7 hours per day with 4 adjustments per year. Using Eq. (21), we find that a circular trough with $\theta_i = 9°$ and $\varphi_{\text{rim}} = 45°$ achieves $C_g = 2.24\times$ which is 43% of the ideal concentration c.f. Eq. (9) . Provided its cost is low, such a concentrator could be useful for photovoltaic or low-temperature solar thermal applications. For example, it could be used to roughly double the power output from a photovoltaic module of a given size, or to increase the temperature and/or thermal efficiency of planar solar thermal collectors, for example the recent

---

[1] Provided that the mirror profile does not cause rays to cross outside the receiver at some intermediate $\varphi$ (which would anyway imply that we are not in the rim ray regime)

designs enabled by transparent aerogel [20, 21], or as an imaging primary coupled to a nonimaging secondary concentrator [22].

In addition to Eq. (21), which enables the overall sizing and position of the receiver for maximum $C_g$, the flux distribution on the receiver may be of interest for a particular application. Although analytical approaches do exist for determining flux distributions [13], Monte Carlo ray tracing (MCRT) is a preferred approach, due to its ability to incorporate detailed effects such as sunshape and slope error. To verify the concentration ratio derivations and further probe the performance of the circular trough, we therefore perform an MCRT[2] simulation to obtain the flux distribution of the abovementioned exemplary design.

**Figure 4** shows the resulting flux distributions for a circular trough and parabolic trough designed for $\theta_i = 9°$ and $\varphi_{rim} = 45°$, which both achieve an identical $C_g$ of 2.24×. All simulations were performed with $10^9$ rays, a direct normal irradiance of 1000 W/m$^2$, and unity reflectivity for the mirror surfaces. Three distributions are shown for each concentrator.

The "filled acceptance cone" was obtained by uniformly illuminating the inlet aperture with radiation filling a cone of half angle equal to the acceptance angle (9°). It is observed that both the circle and parabola achieve essentially the same peak and average flux at the receiver. The zeroing of the flux distribution at the edges of the receiver supports the fulfillment of full collection for both circular and parabolic designs.

The "realistic case, on-axis" includes the effect of sunshape, slope error, and skew dilation. The sunshape was modeled using the correlation from [11] with a circumsolar ratio of 5%, and a normally distributed slope error with standard deviation 3 mrad (acting on the local normal vector) was applied to the mirror. A skew angle of 45° was imposed, which results in an apparent dilation of the solar disk (and sunshape) by a factor equal to sec45° [6]. As seen in **Figure 4**, both the circle and parabola exhibit two peaks in the concentration profile whose location roughly corresponds to the point at which the caustic crosses the receiver (c.f. **Figure 2 b**). The presence of two peaks in the parabola is because, for maximum $C_g$, the receiver is placed slightly below the paraxial focus [18].

---

[2] The Monte Carlo ray-tracing code was developed by the PREC research group of Prof. Steinfeld at ETH Zurich.

The "realistic case, 5° misaligned" uses the same parameters as above, with the additional effect that the concentrator is rotated about its axis by 5° relative to the source. This misalignment represents the effect of either tracking error, or the declination of the sun away from the day when the last seasonal adjustment was made.

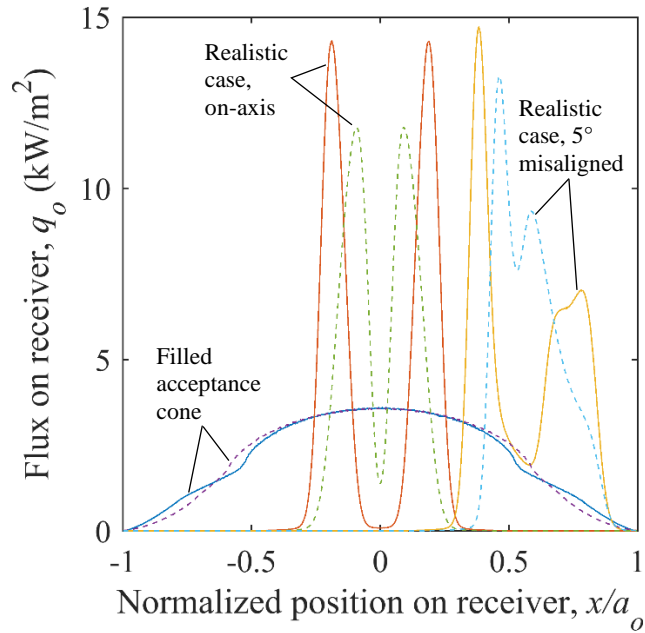

**Figure 4.** Flux distributions for a circular (solid lines) and a parabolic (dashed lines) trough with flat absorber as determined by Monte Carlo ray-tracing ($10^9$ rays). Both concentrators achieve $C_g$ = 2.24× with an acceptance angle of 9°.

Interestingly, for both realistic cases, the circular trough actually achieves a higher peak flux than the parabola. This is likely a disadvantage of the circle, however, as such local hotspots may lead to nonuniform heating for thermal applications, or mismatch losses [23] for photovoltaic applications.

## 5  Conclusion

In this paper, we have presented a simple formula for the maximum full-collection geometric concentration ratio for a circular trough concentrator coupled to a flat receiver as a function of its rim angle and acceptance angle. We have identified the presence of two regimes, one governed by the edge ray caustics, and one governed by the edge rays from the rim alone. It was shown that in the rim ray regime, which is primarily of interest for large acceptance angles designs, the circular trough achieves an identical receiver size (identical $C_g$), and thus essentially matches the performance of a parabolic trough in a nonimaging sense. We performed Monte Carlo ray tracing on an exemplary non-tracking circular trough design which achieves $C_g = 2.24\times$ and allows for year-round operation with seasonal adjustment. Given the relative ease by which a circular profile can be constructed compared to a parabola (e.g. by inflation), the results may be useful for guiding the development of lower cost, lower concentration solar trough designs. The provided $C_g$ formula fills a gap towards developing a complete catalogue of concentration ratio formulas for different concentrator/receiver geometries, for which rigorous formulas for circular concentrators are conspicuously missing. Moreover, the methodology utilized in this paper may well be useful for developing similar $C_g$ formulas for other potentially useful concentrator/receiver geometries. Building on this, in a forthcoming paper we apply a similar methodology to determine the $C_g$ formula of another relevant configuration, namely a circular trough coupled to a *circular* receiver.

**Acknowledgments**

The authors acknowledge the support of the Natural Sciences and Engineering Research Council of Canada (NSERC) through Grant No. RGPIN-2019-06801. We gratefully acknowledge the permission from Prof. Steinfeld (PREC group at ETH Zurich) to use the MCRT code utilized for the ray tracing simulations, and the efforts of previous PREC group members in developing this code.

**Data Availability**

The data that support the findings of this study are available from the corresponding author upon reasonable request.